 \documentclass[prl,twocolumn,showpacs,showkeys,floatfix]{revtex4}

\usepackage{graphicx}% Include figure files
\usepackage{amssymb,amsmath}
\usepackage{dcolumn}% Align table columns on decimal point
\usepackage{bm}% bold math
\usepackage{color}
\usepackage{tabularx}

\begin{document}

\title{Long-range magnetic order and spin-lattice coupling in the 
       delafossite $ {\bf CuFeO_2} $}

\author{Volker Eyert}
\email{eyert@physik.uni-augsburg.de}
\altaffiliation[Permanent address: ]
               {Center for Electronic Correlations and Magnetism,
                Institut f\"ur Physik, Universit\"at Augsburg, 
                86135 Augsburg, Germany}
\affiliation{Laboratoire CRISMAT, UMR CNRS-ENSICAEN(ISMRA) 6508, 
             6, Boulevard Mar\'echal Juin, 14050 Caen Cedex, France}
\author{Raymond Fr\'esard}
\affiliation{Laboratoire CRISMAT, UMR CNRS-ENSICAEN(ISMRA) 6508, 
             6, Boulevard Mar\'echal Juin, 14050 Caen Cedex, France}
\author{Antoine Maignan}
\affiliation{Laboratoire CRISMAT, UMR CNRS-ENSICAEN(ISMRA) 6508, 
             6, Boulevard Mar\'echal Juin, 14050 Caen Cedex, France}

\date{\today}

\begin{abstract}
The electronic and magnetic properties of the delafossite $ {\rm CuFeO_2} $ 
are investigated by means of electronic structure calculations. They are 
performed using density functional theory in the generalized gradient 
approximation as well as the new full-potential augmented spherical wave 
method. The calculations reveal three different spin states at the iron 
sites. Taking into account the correct crystal structure, we find long-range 
antiferromagnetic ordering in agreement with experiment. Contrasting 
previous work, our calculations show that non-local exchange interactions lead
to a semiconducting ground state. 
\end{abstract}



\maketitle

%\section{Introduction}
%\label{intro}

Long-range magnetic ordering on triangular lattices with 
antiferromagnetic exchange interactions is at the focus of continuing 
interest. This is due to the strong geometric frustration experienced 
by such systems, which may lead to a large variety of magnetic states 
including incommensurate and non-collinear spin arrangements. As a 
consequence, there is usually a complex response to magnetic fields, 
which gives rise, e.g., to magnetization steps
\cite{kageyama97,maignan00,hardy06}. While this behavior has been 
observed in the trigonal chain cobaltates, where it leads to a striking 
spin dynamics \cite{maignan00}, the genuine situation is that of well 
separated triangular layers as they are found, e.g., in the 
delafossite-type compounds $ {\rm ABO_2} $. 

In general, this broad class of materials has aroused much interest 
due to a broad range of exciting physical properties \cite{dupont71}, 
including strongly anisotropic very high electrical conductivities 
as, e.g., in $ {\rm PdCoO_2} $ and semiconducting behavior in 
antiferromagnetic delafossites. High optical band gaps in Cu- and 
Ag-based materials allow for simultaneous transparency and p-type 
conductivity \cite{kawazoe97} and, hence, the development of 
transparent optoelectronic devices. This large variety is caused by 
the stacking of monoatomic triangular layers within the rhombohedral 
structure \cite{dupont71}. Edge-sharing distorted oxygen octahedra 
surrounding the B-atoms form $ {\rm BO_2} $ sandwich layers, which 
are linked to the A-atom layers via linear O--A--O bonds (for the 
crystal structure see also Fig.\ 1 of Ref.\ \cite{eyert08}). 
Generically, the A and B atoms are mono- and trivalent, respectively. 
Depending on the chemical composition, this opens a zoo of behaviors: 
for instance, if the $ {\rm A^+} $ ion is in a $ d^9 $ configuration 
good metallic conductivity is observed as in the case of 
$ {\rm PdCoO_2} $, while if it is in a $ d^{10} $ configuration, 
the degrees of freedom dominating the low-energy physics are due 
to the B atoms as, e.g., in $ {\rm CuCrO_2} $ and $ {\rm AgCoO_2} $.  

As for many other magnetic delafossite compounds, the exact magnetic 
structure of $ {\rm CuFeO_2} $ has long been a matter of dispute 
\cite{muir67,doumerc86}. Using neutron diffraction, Mekata {\em et al.}\ 
were able to distinguish two different magnetic phases below 
$ {\rm T_{N1} = 16} $\,K and $ {\rm T_{N2} = 11} $\,K. They are 
connected with monoclinic and orthorhombic magnetic supercells, 
respectively, of the undistorted rhombohedral unit cell with 
commensurate and incommensurate collinear 
arrangements of the localized $ 4.4 \mu_B $ $ {\rm Fe^{3+}} $ moments 
\cite{mekata92,mekata93,petrenko00,kimura06}. Observation of a 
noncollinear-incommensurate phase in magnetic field was taken as 
indicative of possible multiferroic behavior \cite{kimura06}, 
which was indeed observed in Al-doped $ {\rm CuFeO_2} $ \cite{seki07}. 
Quite recently, X-ray and neutron diffraction measurements by 
Ye {\em et al.}\ contrasted the previous observations by revealing 
structural distortions accompanying the magnetic phase transitions, 
which eventually lead to a monoclinic structure at 4\,K \cite{ye06}.

Only few electronic structure calculations for magnetic delafossite 
compounds have been reported in the literature 
\cite{galakhov97,seshadri98,ong07,singh07}. From LDA calculations, 
Galakhov {\em et al.}\ obtained a ferromagnetic state for the 
rhombohedral $ R\bar{3}m $ structure with a magnetic moment at the Fe 
site of about 0.9\, $ \mu_B $, much lower than the experimental value 
\cite{galakhov97}. The Fe $ 3d $ $ t_{2g} $ states were found above 
the Cu $ 3d $ states just at $ {\rm E_F} $, in disagreement with both 
photoemission data and the fact that $ {\rm CuFeO_2} $ 
is a semiconductor with an optical band gap of about 1.15\,eV. In 
contrast, LDA+U calculations led to a band gap of 2\,eV and a magnetic 
moment of 3.76\,$ \mu_B $. However, the occupied Fe $ 3d $ states were 
located at about 9\,eV below the valence band 
maximum and thus much to low \cite{galakhov97}. Recent calculations by 
Ong {\em et al.}\ using the generalized gradient approximation (GGA) 
\cite{pbe96} resulted in a high-spin state with a magnetic moment of 
3.78\,$ \mu_B $ per Fe and the Fe $ 3d $ $ t_{2g} $ spin-up states 
below the Cu $ 3d $ bands in agreement with photoemission and X-ray 
emission data \cite{ong07}. However, again a finite optical band gap 
was arrived at only after taking into account electronic correlations 
within the LDA+U scheme \cite{ong07}. 

In the present work we apply the new full-potential augmented spherical 
wave method to study the electronic properties of $ {\rm CuFeO_2} $. In 
doing so, we take for the first time the experimentally observed 
low-temperature structure into account. Our calculations resolve open 
issues by revealing i) an antiferromagnetic ground state for the 
monoclinic structure in perfect agreement with the experimental data, 
ii) the opening of a fundamental band gap already at the GGA level, 
i.e.\ without taking local correlations into account, and 
iii) the quite unusual existence of three different magnetic states of 
assumed ferromagnetic $ {\rm CuFeO_2} $, which so far has been seen 
only for elemental iron \cite{moruzzi86}.

%\section{Theoretical Method}
%\label{method}

The calculations are based on density-functional theory and the 
generalized gradient approximation (GGA) \cite{pbe96} with the
local-density approximation parameterized according to Perdew and
Wang \cite{perdew92}. They were performed using the scalar-relativistic 
implementation of the augmented spherical wave (ASW) method (see Refs.\ 
\onlinecite{wkg,aswrev,aswbook} and references therein).
In the ASW method, the wave function is expanded in atom-centered
augmented spherical waves, which are Hankel functions and numerical
solutions of Schr\"odinger's equation, respectively, outside and inside
the so-called augmentation spheres. In order to optimize the basis set,
additional augmented spherical waves were placed at carefully selected
interstitial sites. The choice of these sites as well as the augmentation
radii were automatically determined using the sphere-geometry optimization
algorithm \cite{sgo}. Self-consistency was achieved by a highly efficient
algorithm for convergence acceleration \cite{mixpap}. The Brillouin zone
integrations were performed using the linear tetrahedron method with up
to 1156 and 3180 {\bf k}-points within the irreducible wedge of the 
rhombohedral and monoclinic Brillouin zone, respectively 
\cite{bloechl94,aswbook}. For the monoclinic magnetic supercell, up to 
100 {\bf k}-points were used. 

In the present work, a new full-potential version of the ASW method 
was employed \cite{fpasw}. 
In this version, the electron density and related quantities are given
by spherical-harmonics expansions inside the muffin-tin spheres.
In the remaining interstitial region, a representation in terms of
atom-centered Hankel functions is used \cite{msm88}. However, in
contrast to previous related implementations, we here get away without
needing a so-called multiple-$ \kappa $ basis set, which fact allows to 
investigate rather large systems with a minimal effort.

%\section{Results and Discussion}
%\label{results}

\begin{figure}[htb]
\centering
\includegraphics[width=\columnwidth,clip]{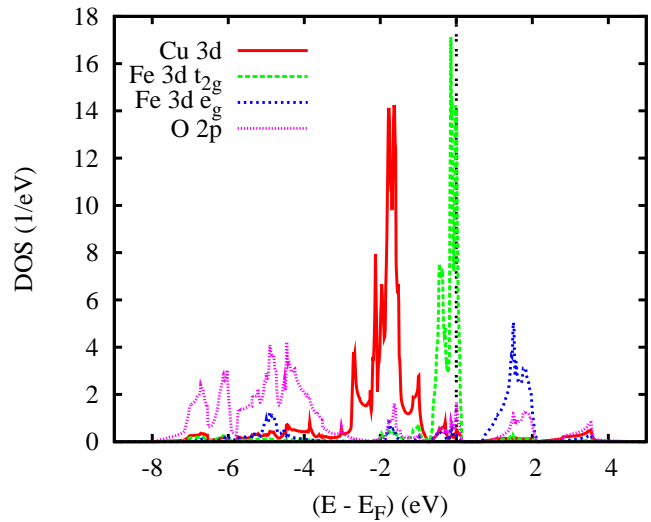}
\vspace*{-1em}
\caption{(Color online) Partial densities of states (DOS) of 
         rhombohedral $ {\rm CuFeO_2} $.
         Selection of Fe $ 3d $ orbitals in this and the subsequent 
         figures is relative to the local rotated reference frame, 
         see text.}
\vspace*{-1em}
\label{fig1}
\end{figure}

The calculations used the crystal structure data by Ye {\em et al.}\ 
\cite{ye06}. As a starting point, spin-degenerate calculations for the 
rhombohedral structure were performed. The resulting partial densities 
of states (DOS) are shown in Fig.\ \ref{fig1}.
While the lower part of the spectrum is dominated by O $ 2p $ states, 
the transition metal $ d $ states lead to rather sharp peaks in the 
interval from $ -3 $ to $ +2 $\,eV. In particular, the $ t_{2g} $ 
and $ e_g $ manifolds of the Fe $ 3d $ states as resulting from the 
octahedral coordination are recognized. This representation of the 
partial DOS used a local rotated coordinate system with the Cartesian 
axes pointing towards the oxygen atoms. $ \sigma $-type overlap of the 
O $ 2p $ states with the Fe $ 3d $ $ e_g $ orbitals leads to the 
contribution of the latter near $ -5 $\,eV. In contrast, due to the 
much weaker $ \pi $-type overlap of the O $ 2p $ states with the 
$ t_{2g} $ orbitals, these states give rise to sharp peaks in the 
interval from $ -0.8 $\,eV to just above the Fermi energy. The latter 
falls right into the upper part of the $ t_{2g} $ manifold and Fe 
turns out to be in a $ d^5 $ state. 
In contrast, the Cu $ 3d $ states are essentially limited to the 
interval from $ -3 $ to $ -1 $\,eV and thus Cu can be assigned a 
monovalent $ d^{10} $ configuration in close analogy with the 
experimental findings. In passing, we mention the finite dispersion 
of the electronic bands parallel to $ \Gamma $-A, which points to a 
considerable three-dimensionality arising from the coupling between 
the layers as has been observed also in other delafossite materials 
\cite{eyert08}. 

\begin{figure}[htbp!]
\centering
\includegraphics[width=\columnwidth]{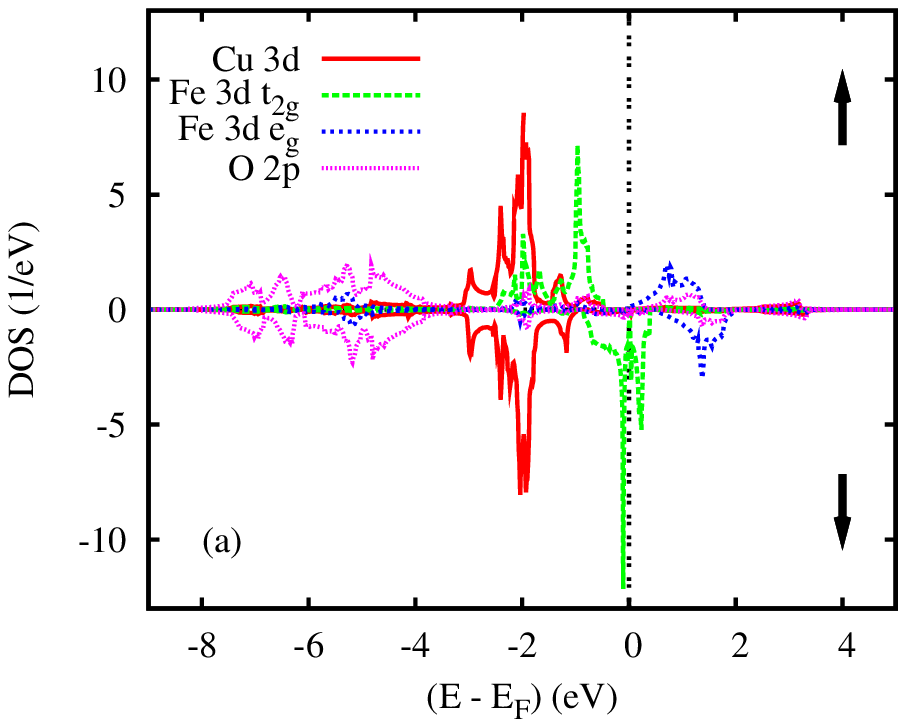}
\includegraphics[width=\columnwidth]{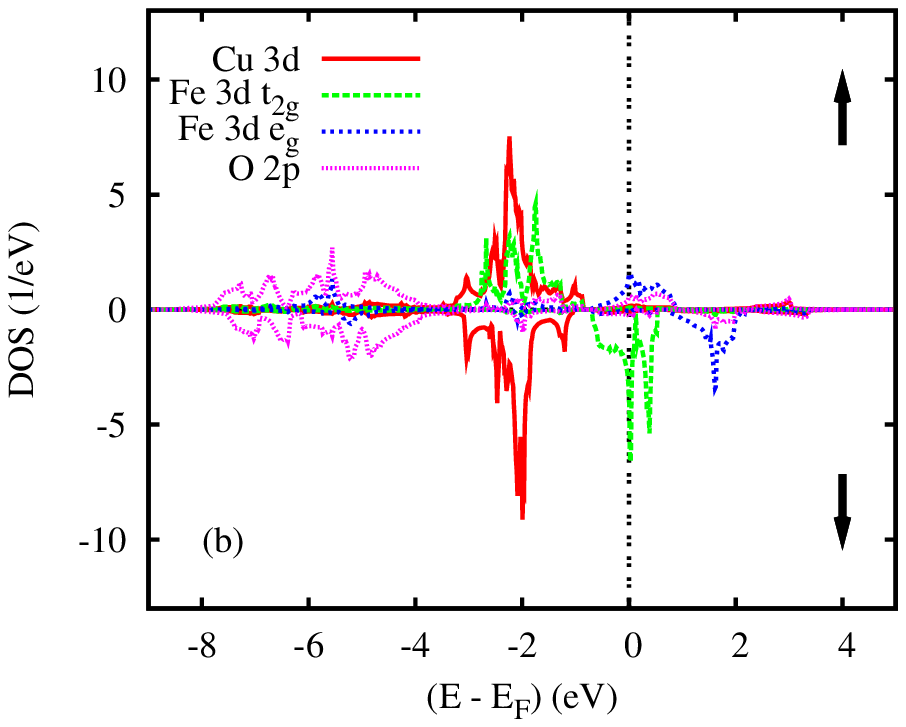}
\includegraphics[width=\columnwidth]{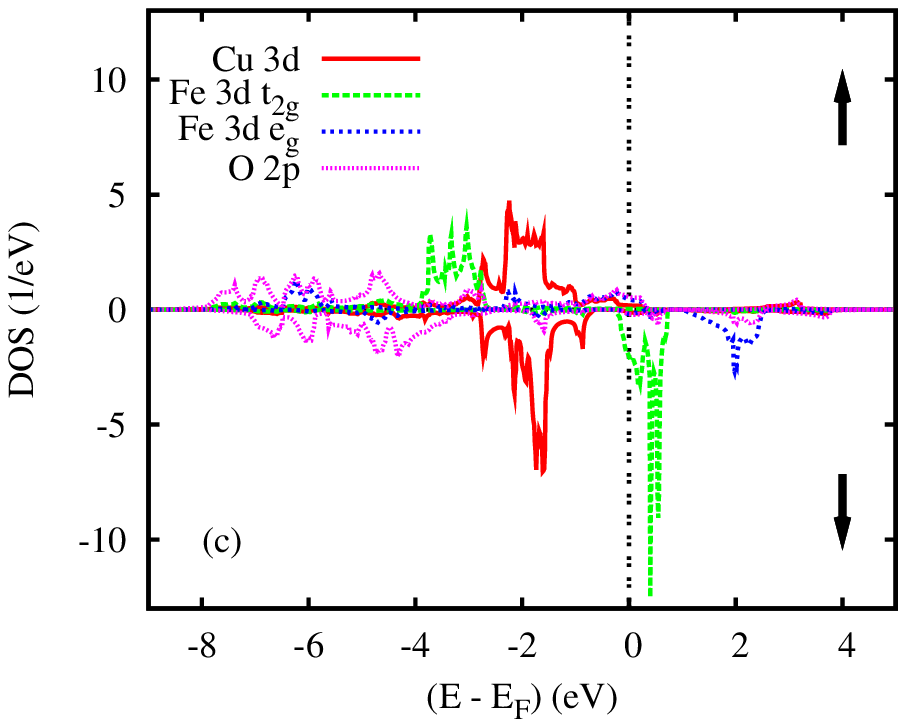}
\caption{(Color online) Partial densities of states (DOS) of 
         rhombohedral ferromagnetic (a) low-spin, (b) intermediate-spin, 
         and (c) high-spin $ {\rm CuFeO_2} $.}
\label{fig2}
\end{figure}

Since the perfect triangular lattice of the rhombohedral structure 
does not allow for long-range antiferromagnetic order, subsequent 
spin-polarized calculations were performed for an assumed ferromagnetic 
state in a  spirit similar to the previous work by Galakhov 
{\em et al.}\ as well as by Ong {\em et al.}\ \cite{galakhov97,ong07}. 
From our calculations, three different configurations were obtained 
corresponding to a low-spin, intermediate-spin, and high-spin moment 
located at the Fe site. The respective partial densities of states 
are displayed in Fig.\ \ref{fig2}.
The total energies as compared to the spin-degenerate configuration 
and the local magnetic moments are summarized in Tab.\ \ref{tab1}.
\begin{table}
\caption{Total energies (in mRyd per formula unit) and magnetic 
         moments (in $ \mu_B $) for different crystal structures 
         and magnetic orderings of $ {\rm CuFeO_2} $.}
\begin{ruledtabular}
\begin{tabular}{ccrrr} 
structure  & magn.\ order & $ \Delta E $ & $ m_{\rm Fe} $ & $ m_{\rm O} $ \\
\hline
rhomb.\    & spin-deg.\   & $      0.0 $ &                &               \\
rhomb.\    & ferro (LS)   & $    -16.7 $ & $       1.03 $ & $    - 0.02 $ \\
rhomb.\    & ferro (IS)   & $    -12.0 $ & $       2.02 $ & $    - 0.02 $ \\
rhomb.\    & ferro (HS)   & $    -19.2 $ & $       3.73 $ & $      0.21 $ \\
monoclinic & spin-deg.\   & $     -6.0 $ &                &               \\
monoclinic & ferro (LS)   & $    -21.5 $ & $       1.04 $ & $    - 0.02 $ \\
monoclinic & ferro (IS)   & $    -19.0 $ & $       2.08 $ & $    - 0.02 $ \\
monoclinic & ferro (HS)   & $    -32.0 $ & $       3.62 $ & $      0.19 $ \\
monoclinic & antiferro    & $    -46.0 $ & $   \pm 3.72 $ & $  \pm 0.08 $ \\
\end{tabular}
\end{ruledtabular}
\label{tab1}
\end{table}
In general, the observation of three different spin states for magnetic 
ions is very unusual and so far has been reported only for elemental 
fcc Fe \cite{moruzzi86}. 
According to the partial DOS, the magnetic moments of the low-spin and 
intermediate-spin states are almost exclusively carried by the Fe $ 3d $ 
$ t_{2g} $ states, which show a spin splitting of about $ 1 $ and 
$ 2 $\,eV, respectively. In the high-spin configuration this splitting 
increases to $ \approx 3.5 $\,eV and the magnetic moment is carried by 
both the $ t_{2g} $ and $ e_g $ states. In addition, due to the strong 
$ \sigma $-type overlap with the latter, the O $ 2p $ states 
also experience a substantial polarization. For the same reason, the 
$ e_g $ moments start to form already in the energy interval of the 
O $ 2p $ states leading to distinctly different spin-up and spin-down 
$ e_g $ partial DOS. A similar behavior has been also observed in other 
high-spin systems and termed as the formation of local extended magnetic 
moments \cite{eyert04}. 
As is obvious from Tab.\ \ref{tab1}, all three ferromagnetic configurations  
have energies lower than the spin-degenerate situation. However, the 
high-spin state is most stable for the rhombohedral lattice. 
In summary, our calculations not only reproduce both the low-spin and 
high-spin results obtained by Galakhov {\em et al.}\ as well as by Ong 
{\em et al.}\ and explain the differences between their findings but 
additionally prove the existence of yet another, intermediate-spin 
state. Furthermore, the high-spin partial DOS compare very well with 
the photoemission and X-ray emission data \cite{galakhov97}. 

In a second step, the monoclinic structure observed by Ye {\em et al.}\ 
was considered \cite{ye06}. Note that in this structure there is still 
only one Fe atom per unit cell. 
Both the spin-degenerate and the spin-polarized ferromagnetic calculations 
led to essentially the same partial DOS as for the rhombohedral structure. 
In particular, again three different magnetic configurations were found 
with the local magnetic moments as listed in Tab.\ \ref{tab1} being almost 
identical to those obtained for the rhombohedral structure. However, the 
total energies, also given in Tab.\ \ref{tab1}, are generally lower by 
several mRyd with the largest energy lowering occurring for the high-spin 
state. 

\begin{figure}[htb]
\centering
\includegraphics[width=\columnwidth,clip]{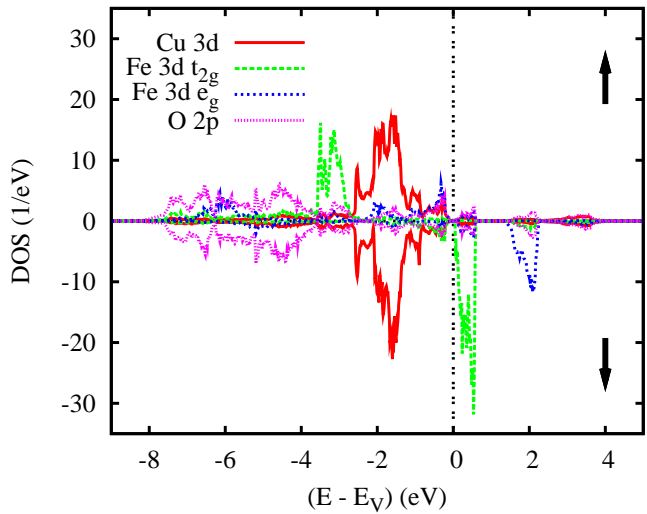}
\vspace*{-1em}
\caption{(Color online) Partial densities of states (DOS) of 
         monoclinic antiferromagnetic high-spin $ {\rm CuFeO_2} $.}
\vspace*{-1em}
\label{fig5}
\end{figure}

Finally, calculations for the eightfold magnetic supercell proposed by 
Ye {\em et al.} \cite{ye06} were performed. The resulting partial DOS 
are displayed in Fig.\ \ref{fig5} 
and the local magnetic moments and total energy included in Tab.\ 
\ref{tab1}. According to these results, the antiferromagnetic state 
has the lowest energy as compared to all other configurations. In 
addition, Fe is found to be in a high-spin state in agreement with 
the neutron diffraction data by Mekata {\em et al.}\ 
\cite{mekata92,mekata93}. Remarkably, a band gap of 0.05\,eV is 
obtained. Thus the non-local exchange interaction included in the 
GGA leads to the semiconducting ground state once the monoclinic 
structure is correctly accounted for. However, the band gap is too 
small reflecting the well known shortcomings of the GGA. Additional 
inclusion of electronic correlations, for instance via the LDA+U method, 
is needed to achieve quantitative agreement with experiment. 

%\section{Summary}
%\label{summary}

In summary, our calculations for $ {\rm CuFeO_2} $ demonstrate that 
i) taking the monoclinic structure into account results in an 
antiferromagnetic ground state in perfect agreement with the experimental 
situation, ii) a fundamental band gap is opened already within the GGA, 
and iii) there is a quite unusual competition among several magnetic states, 
including {\em three} different magnetic states of assumed ferromagnetic 
$ {\rm CuFeO_2} $, which so far seems not to have been obtained for iron 
compounds. Concerning the latter point it is remarkable that the trigonal 
environment of  ${\rm Fe^{3+}} $ renders its magnetic states very close 
in energy, while under most circumstances  ${\rm Fe^{3+}} $ is in the 
high spin state. Even though the effect is less pronounced than in the 
calcium cobaltates, where the environment has a dramatic effect on the 
spin configuration of the  ${\rm Co^{3+}} $ ions \cite{aasland97,eyert04}, 
this opens a route to the observation of spin state transitions in 
$ {\rm Fe^{3+}} $ ions as well.

\section{Acknowledgements}
We gratefully acknowledge many useful discussions with T.\ Kopp, 
C.\ Martin, and W.\ C.\ Sheets. 
This work was supported by the Deutsche Forschungsgemeinschaft through 
SFB 484.

%
%  References
%


\begin{thebibliography}{}

\bibitem{kageyama97}
H.\ Kageyama, K.\ Yoshimura, K.\ Kosuge, H.\ Mitamura, and T.\ Goto,
J.\ Phys.\ Soc.\ Japan {\bf 66},  1607 (1997).

\bibitem{maignan00}
A.\ Maignan, C.\ Michel, A.\ C.\ Masset, C.\ Martin, and B.\ Raveau,
Eur.\ Phys.\ J.\ B {\bf 15},  657 (2000).

\bibitem{hardy06}
V.\ Hardy, C.\ Martin, G.\ Martinet, and G.\ Andr\'e,
Phys.\ Rev.\ B {\bf 74}, 064413 (2006).

\bibitem{dupont71} 
%\bibitem{shannon71} 
R.\ D.\ Shannon, D.\ B.\ Rogers, and C.\ T.\ Prewitt, 
Inorg.\ Chem.\ {\bf 10}, 713 (1971); 
%\bibitem{prewitt71} 
C.\ T.\ Prewitt, R.\ D.\ Shannon, and D.\ B.\ Rogers, 
{\em ibid.}\ {\bf 10}, 719 (1971); 
%\bibitem{rogers71} 
D.\ B.\ Rogers, R.\ D.\ Shannon, C.\ T.\ Prewitt, and J.\ L.\ Gillson, 
{\em ibid.}\ {\bf 10}, 723 (1971). 

\bibitem{kawazoe97} 
H.\ Kawazoe, M.\ Yasukawa, H.\ Hyodo, M.\ Kurita, H.\ Yanagi, and H.\ Hosono, 
Nature {\bf 389}, 939 (1997). 

\bibitem{eyert08} 
V.\ Eyert, R.\ Fr\'esard, and A.\ Maignan,
arXiv:0801.4077 (unpublished)

\bibitem{muir67} 
A.\ H.\ Muir and M.\ Wiedersich, 
J.\ Phys.\ Chem.\ Solids {\bf 28}, 65 (1967). 

\bibitem{doumerc86} 
J.-P.\ Doumerc, A.\ Wichainchai, A.\ Ammar, M.\ Pouchard, and P.\ Hagenmuller, 
Mat.\ Res.\ Bull.\ {\bf 21}, 745 (1986). 

\bibitem{mekata92} 
M.\ Mekata, N.\ Yaguchi, T.\ Takagi, S.\ Mitsuda, and H.\ Yoshizawa, 
J.\ Magn.\ Magn.\ Mater.\ {\bf 104-107}, 823 (1992). 

\bibitem{mekata93} 
M.\ Mekata, N.\ Yaguchi, T.\ Takagi, T.\ Sugino, S.\ Mitsuda, H.\ Yoshizawa, 
N.\ Hosoito, and T.\ Shinjo, 
J.\ Phys.\ Soc.\ Japan {\bf 62}, 4474 (1993). 

\bibitem{petrenko00}
O.\ A.\ Petrenko, G.\ Balakrishnan, M.\ R.\ Lees, D.\ McK.\ Paul, and 
A.\ Hoser. 
Phys.\ Rev.\ B {\bf 62}, 8983 (2000).

\bibitem{kimura06}
T.\ Kimura, J.\ C.\ Lashley, and A.\ P.\ Ramirez, 
Phys.\ Rev.\ B {\bf 73}, 220401(R) (2006).

\bibitem{seki07}
S.\ Seki, Y.\ Yamasaki, Y.\ Shiomi, S.\ Iguchi, Y.\ Onose, and Y.\ Tokura, 
Phys.\ Rev.\ B {\bf 75}, 100403(R) (2007).

\bibitem{ye06}
F.\ Ye, Y.\ Ren, Q.\ Huang, J.\ A.\ Fernandez-Baca, P.\ Dai, J.\ W.\ Lynn, 
and T.\ Kimura, 
Phys.\ Rev.\ B {\bf 73}, 220404(R) (2006).

\bibitem{galakhov97}
V.\ R.\ Galakhov, A.\ I.\ Poteryaev, E.\ Z.\ Kurmaev, V.\ I.\ Anisimov, 
S.\ Bartkowski, M.\ Neumann, Z.\ W.\ Lu, B.\ M.\ Klein, and T.-R.\ Zhao, 
Phys.\ Rev.\ B {\bf 56}, 4584 (1997);

\bibitem{seshadri98}
R.\ Seshadri, C.\ Felser, K.\ Thieme, and W.\ Tremel, 
Chem.\ Mater.\ {\bf 10}, 2189 (1998);

\bibitem{ong07} 
K.\ P.\ Ong, K.\ Bai, P.\ Blaha, and P.\ Wu, 
Chem.\ Mater.\ {\bf 19}, 634 (2007); 

\bibitem{singh07}
D.\ J.\ Singh, 
Phys.\ Rev.\ B {\bf 76}, 085110 (2007).

\bibitem{moruzzi86}
V.\ L.\ Moruzzi, P.\ M.\ Marcus, K.\ Schwarz, and P.\ Mohn, 
Phys.\ Rev.\ B {\bf 34}, 1784 (1986).

\bibitem{pbe96}
J.\ P.\ Perdew, K.\ Burke, and M.\ Ernzerhof,
Phys.\ Rev.\ Lett.\ {\bf 77}, 3865 (1996).

\bibitem{perdew92}
J.\ P.\ Perdew and Y.\ Wang,
Phys.\ Rev.\ B {\bf 45}, 13244 (1992).

\bibitem{wkg}
A.\ R.\ Williams, J.\ K\"ubler, and C.\ D.\ Gelatt, Jr., 
Phys.\ Rev.\ B {\bf 19}, 6094 (1979).

\bibitem{aswrev}
V.\ Eyert, 
Int.\ J.\ Quantum Chem.\, {\bf 77}, 1007 (2000).

\bibitem{aswbook}
V.\ Eyert,
{\em The Augmented Spherical Wave Method -- A Comprehensive Treatment},
Lect.\ Notes Phys.\ {\bf 719} (Springer, Berlin Heidelberg 2007).

\bibitem{sgo}
V.\ Eyert and K.-H.\ H\"ock, 
Phys.\ Rev.\ B {\bf 57}, 12727 (1998).

\bibitem{mixpap}
V.\ Eyert, 
J.\ Comp.\ Phys.\ {\bf 124}, 271 (1996).

\bibitem{bloechl94}
P.\ E.\ Bl\"ochl, O.\ Jepsen, and O.\ K.\ Andersen,
Phys.\ Rev.\ B {\bf 49}, 16223 (1994).

\bibitem{fpasw}
V.\ Eyert, 
J.\ Comp.\ Chem., in press. 

\bibitem{msm88}
M.\ S.\ Methfessel, 
Phys.\ Rev.\ B {\bf 38}, 1537 (1988).

\bibitem{eyert04}
V.\ Eyert, C.\ Laschinger, T.\ Kopp, and R.\ Fr\'{e}sard, 
Chem.\ Phys.\ Lett.\ {\bf 385}, 249 (2004).

\bibitem{aasland97}
S.\ Aasland, H.\ Fjellv\aa g, and B.\ C.\ Hauback,
Solid State Comm.\ {\bf 101}, (1997) 187.

\end{thebibliography}
\end{document}